\documentclass{mnras}
\usepackage{graphicx}

\newcommand\msun{{\rm M_{\odot}}}
\newcommand\lsun{{\rm L_{\odot}}}

\def\go{
\mathrel{\raise.3ex\hbox{$>$}\mkern-14mu\lower0.6ex\hbox{$\sim$}}
}

\title[V745 Sco]
{{\em Swift} detection of the super-swift switch-on of the super-soft phase in nova V745 Sco (2014)}
\author[K.L. Page et al.]{K.L. Page$^{1}$, J.P. Osborne$^{1}$, N.P.M. Kuin$^{2}$, M. Henze$^{3}$, F.M. Walter${^4}$,  \newauthor A.P. Beardmore$^{1}$, M.F. Bode$^{5}$, M.J. Darnley$^{5}$, L. Delgado$^{3}$, J.J. Drake$^6$, \newauthor M. Hernanz$^{3}$,  K. Mukai$^{7,8}$, T. Nelson$^{9}$, J.-U. Ness$^{10}$, G.J. Schwarz$^{11}$, \newauthor S.N. Shore$^{12,13}$,  S. Starrfield$^{14}$ and C.E. Woodward$^{9}$\\
$^{1}$ X-Ray and Observational Astronomy Group, Department of Physics \&
  Astronomy, University of Leicester, LE1 7RH, UK\\
$^{2}$ Mullard Space Science Laboratory/University College London, Holmbury St. Mary, Dorking, Surrey, RH5 6NT, UK\\
$^{3}$ Institut de Ci{\` e}ncies de l'Espai (CSIC-IEEC), Campus UAB, C/Can Magrans s/n, E-08193 Cerdanyola del Valles, Spain\\
$^{4}$ Department of Physics \& Astronomy, Stony Brook University, Stony Brook, NY 11794-3800, USA\\
$^{5}$ Astrophysics Research Institute, Liverpool John Moores University, IC2 Liverpool Science Park, Liverpool, L3 5RF, UK\\
$^{6}$ Smithsonian Astrophysical Observatory, MS-3, 60 Garden Street, Cambridge, MA 02138, USA\\
$^{7}$ CRESST and X-ray Astrophysics Laboratory, NASA/GCFC, Greenbelt, MD 20771, USA\\
$^{8}$ Department of Physics, University of Maryland, Baltimore, MD 21250, USA\\
$^{9}$  Minnesota Institute of Astrophysics, University of Minnesota, Minneapolis, MN 55455, USA\\
$^{10}$ Science Operations Division, Science Operations Department of ESA, ESAC, Villanueva de la Ca{\~ n}ada, Madrid, Spain\\
$^{11}$ American Astronomical Society, 2000 Florida Ave., NW, Suite 300, DC 20009-1231, USA\\
$^{12}$ Dipartimento di Fisica "Enrico Fermi", Universit{\` a} di Pisa, 56127, Pisa, Italy\\
$^{13}$ INFN-Sezione Pisa, Largo B. Pontecorvo 3, 56127, Pisa, Italy\\
$^{14}$ School of Earth and Space Exploration, Arizona State University, Tempe, AZ 85287, USA\\
}

\date{Accepted XXX. Received YYY; in original form ZZZ}

\pubyear{2015}

\begin{document}
\label{firstpage}
\pagerange{\pageref{firstpage}--\pageref{lastpage}}

\maketitle

\begin{abstract}

V745 Sco is a recurrent nova, with the most recent eruption occurring in February 2014. V745~Sco was first observed by {\em Swift} a mere 3.7~hr after the announcement of the optical discovery, with the super-soft X-ray emission being detected around four days later and lasting for only $\sim$~two days, making it both the fastest follow-up of a nova by {\em Swift} and the earliest switch-on of super-soft emission yet detected.
Such an early switch-on time suggests a combination of a very high velocity outflow and low ejected mass and, together with the high effective temperature reached by the super-soft emission, a high mass white dwarf ($>$1.3~$\msun$).
The X-ray spectral evolution was followed from an early epoch where shocked emission was evident, through the entirety of the super-soft phase, showing evolving column density, emission lines, absorption edges and thermal continuum temperature. UV grism data were also obtained throughout the super-soft interval, with the spectra showing mainly emission lines from lower ionization transitions and the Balmer continuum in emission. V745~Sco is compared with both V2491~Cyg (another nova with a very short super-soft phase) and M31N~2008-12a (the most rapidly recurring nova yet discovered). The longer recurrence time compared to M31N~2008-12a could be due to a lower mass accretion rate, although inclination of the system may also play a part. Nova V745 Sco (2014) revealed the fastest evolving super-soft source phase yet discovered, providing a detailed and informative dataset for study.

\end{abstract}


\section{Introduction}
\label{intro}

Novae are thermonuclear explosions arising in interacting binary systems. Material is transferred from the secondary star onto the white dwarf (WD) primary until the pressure and temperature at the base of the accreted envelope are sufficient to trigger a thermonuclear runaway (TNR; see Bode \& Evans 2008 for a review). Following this initial explosion, the WD surface is typically obscured from view by the ejected material. The ejecta expand, becoming optically thin and often allowing the surface nuclear burning to become visible. As the nuclear-burning-induced wind from the WD declines, the pseudo-photosphere contracts, rising in temperature as it does so. This nuclear-burning emission peaks in the soft X-ray band, and is known as the Super-Soft Source (SSS) state (Krautter 2008).  Eventually nuclear burning can no longer be sustained, and the nova returns to quiescence.

While most novae have only been detected in a single outburst, and are called classical novae (CNe), there are some systems which have shown multiple historic eruptions (as opposed to repeated periods of rebrightening within a specific eruption): these are known as recurrent novae (RNe; see Webbink et al. 1987, Schaefer 2010 and Anupama \& Kamath 2012 for reviews). It is thought that RNe likely have higher WD masses than CNe, together with a higher accretion rate (Starrfield 1989), although T~Pyx, the prototypical (albeit unusual) RN, appears to contain a relatively low mass ($\sim$~1~$\msun$) WD (e.g. Tofflemire et al. 2013; Nelson et al. 2014; Chomiuk et al. 2014).
The secondary stars in RNe are evolved, in comparison to the main sequence secondaries in CNe systems; the high accretion rates are related to the expansion of the donor star as it evolves (for sub-giant RN systems), or to accretion from a wind (in the case of a red giant -- RG -- secondary).
Because of the shorter timescales and higher WD masses involved, RNe accrete, and subsequently eject, smaller amounts of material during each nova cycle (e.g., Wolf et al. 2013). With less ejected matter needed to disperse, the SSS phase of the nova can become visible more rapidly in RNe.

V745~Sco is an RS Oph-like symbiotic system, in which the secondary star is an RG (Duerbeck, Schwarz \& Augusteijn 1989; Sekiuchi et al. 1990; Harrison, Johnson \& Spyromilio 1993). There is some disagreement about the orbital period: Schaefer (2009, 2010) reported it to be 510~days, while Mr{\' o}z et al. (2014) do not confirm this measurement, finding semi-regular pulsations of the RG with periods of 136.5 and 77.4 days. The SMARTS\footnote{http://www.astro.yale.edu/smarts/} (Small and Moderate Aperture Research Telescope System) $I$ and $R$-band photometry for the 2014 nova outburst is consistent with periods of either 77 or 155 days (F.M. Walter, priv. comm.).
V745~Sco had previously been detected in outburst in 1937 (Plaut 1958) and 1989 (Schaefer 2010), with the International Ultraviolet Explorer (IUE) observing the 1989 eruption\footnote{Spectra available from http://ines.ts.astro.it/ines/}. This latest outburst in 2014 adds weight to the suggestion by Schaefer (2010) that there may have been an additional unobserved nova explosion around 1963, giving a recurrence duty cycle of $\sim$~25 years. Schaefer (2010) gives a distance estimate to V745~Sco of (7.8~$\pm$~1.8)~kpc.

The most recent outburst of V745~Sco was reported by R. Stubbings in AAVSO (American Association of Variable Star Observers) Special Notice 380\footnote{http://www.aavso.org/aavso-special-notice-380} (see also Waagen 2014), at a magnitude of 9.0. Throughout this paper, this discovery date of 2014 February 6.694 UT is taken as T$_{0}$. A previous observation by Stubbings 24 hours earlier showed no evidence for the nova, with the star fainter than 13.0 mag.

Following the announcement of its recurrence, V745~Sco was observed across the electromagnetic spectrum, from radio wavelengths to $\gamma$-rays. Rupen et al. (2014) detected rising radio emission within two days of the optical nova, consistent with either optically thick thermal or self-absorbed synchrotron emission, while Kantharia et al. (2014) presented a later rise ($\sim$~26 days after outburst) in the synchrotron emission at longer radio wavelengths. 
Banerjee et al. (2014a,b) reported near-infrared observations, finding large outflow velocities ($\sim$~4000~km~s$^{-1}$ Full Width Half Maximum, or $>$~9000~km~s$^{-1}$ Full Width at Zero Intensity, for Pa~$\beta$), with Anupama et al. (2014) measuring similar line profiles in the optical band. 
Mr{\' o}z et al. (2014) presented observations obtained by the Optical Gravitational Lensing Experiment (OGLE) for many years before, as well as during, the eruption.
Preliminary reports on the {\em Swift} X-ray data, both the early hard and later super-soft emission, were presented by Mukai et al. (2014), Page et al. (2014a,b) and Beardmore, Osborne \& Page (2014). {\em NuSTAR} (Rana et al. 2014; Orio et al. 2015) observed V745~Sco ten days after outburst, after the SSS had peaked and was starting to fade, while the {\em Chandra} observation (Drake et al. 2014), triggered by the {\em Fermi}-LAT (Large Area Telescope; Atwood et al. 2009) detection (see below), occurred over days 16--17, after the SSS phase had ended. Shocked emission lines were still detected by {\em Chandra} at this time. In addition, Luna et al. (2014) analysed {\em XMM-Newton} data of V745~Sco in quiescence in 2010, finding a weak X-ray source.

Cheung, Jean \& Shore (2014a) reported a {\em Fermi}-LAT detection of V745~Sco, with high energy emission being found (at 2--3$\sigma$) on 2014 February 6 and 7 (the day of outburst and the following day). V745~Sco is, thus, the sixth $\gamma$-ray nova detected by the LAT, after V407~Cyg, V1324~Sco, V959~Mon, V339~Del and V1369~Cen (Cheung et al. 2014b; Cheung, Jean \& Shore 2013), albeit not a very strong detection. The  {\em Swift}-BAT (Burst Alert Telescope; Barthelmy et al. 2005) Transient Monitor (Krimm et al. 2013) found no significant detection over 15--50~keV (H.A. Krimm, priv. comm.).

Here we present {\em Swift} (Gehrels et al. 2004) and SMARTS observations of V745~Sco, showing an extremely rapid rise and fall of the SSS X-ray emission. Errors are given at the 90~per~cent confidence level, unless otherwise stated. The abundances from Wilms, Allen \& McCray (2000) and photoelectric absorption cross-sections from Verner et al. (1996) have been assumed for the X-ray spectral modelling.

\section{Observations}
\label{obs}

\begin{figure*}
\begin{center}
\includegraphics[clip, angle=-90, width=16cm]{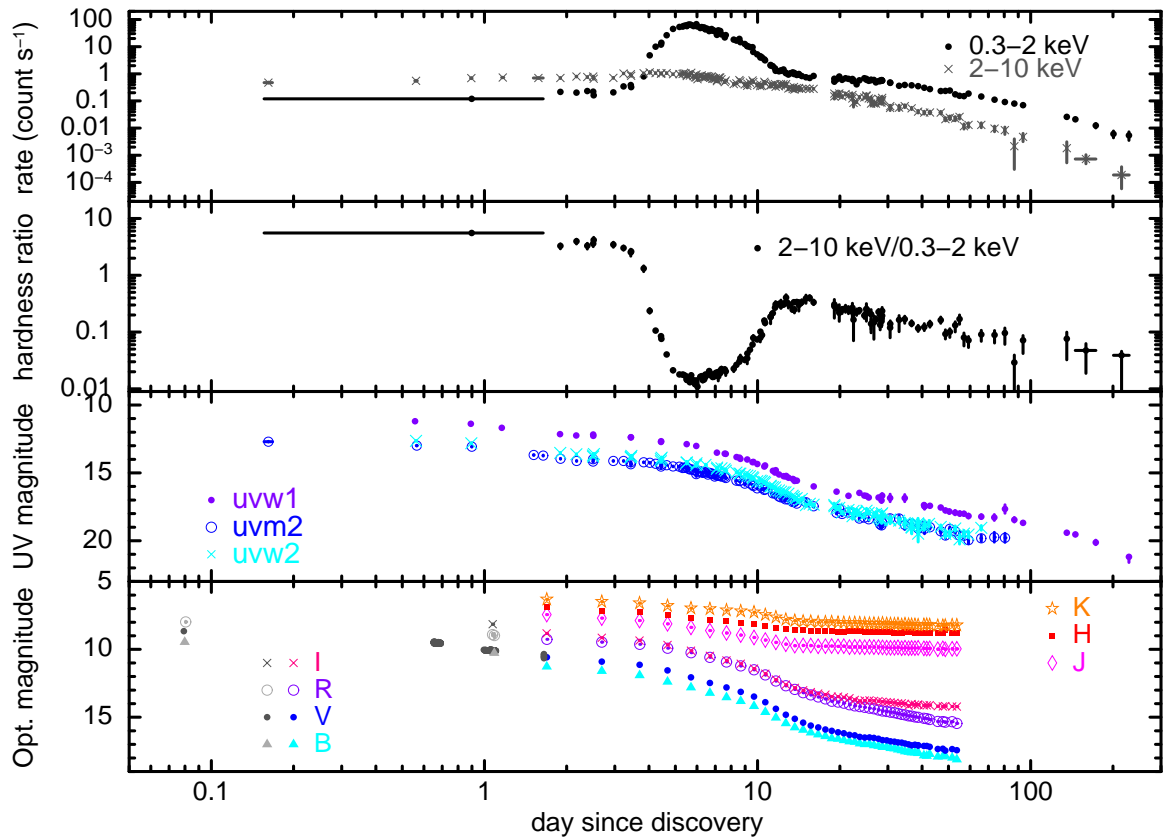}
\caption{{\em Swift} X-ray soft and hard-band light-curves (top panel), hardness ratio (second panel), {\em Swift} UV light-curves (third panel) and {\em SMARTS} and AAVSO optical/IR light-curves (bottom panel) of V745 Sco. The AAVSO data points are those earlier than day 1.1, shown by grey-scale markers.}
\label{lc}
\end{center}
\end{figure*}

\subsection{Swift}

{\em Swift} began observations of V745~Sco only 3.7~hr after the optical discovery, making it the fastest follow-up of a nova by {\em Swift} to date. An initial monitoring campaign of XRT (X-ray Telescope; Burrows et al. 2005) and UVOT (UV/Optical Telescope; Roming et al. 2005) observations every $\sim$~6~hr was begun immediately, followed by multiple snapshots of data being obtained most days until 2014 March 6 (28 days after discovery), by which time the SSS phase was over. Regular observations were continued, though with a decreasing cadence as the X-ray and UV source faded, until the end of 2014 September. 

Initially V745~Sco was observed using all three UV filters ($uvw1$ with a central wavelength of 2600~\AA; $uvm2$ -- 2246~\AA; $uvw2$ -- 1928~\AA); the $u$-band filter (3465~\AA) was also used until day 2.17, but those data suffered from significant coincidence loss due to the brightness of the source. By day 90, the UV source had faded sufficiently such that it was only detectable in the  $uvw1$ filter, so observations using the other filters ceased. By the end of the observing window in 2014 September, the nova was barely detectable with the $uvw1$ filter either. The {\em Swift} UVOT photometry was analysed using the {\sc uvotsource} tool, with the Vega zero points from Breeveld et al. (2011).

In addition, {\em Swift} started collecting UV grism data (covering a wavelength range of 1700--5000 \AA; Kuin et al. 2015) on day 1.16, obtaining 35 spectra before day 38.71, although the nova was too faint after day 15.6 to extract useful information from the grism. These observations were all made with the UV grism in clocked mode, using 
an offset to reduce contamination of zeroth orders in the very crowded field.


The {\em Swift} data were processed and analysed using HEASoft version 6.16 with the most recent (pre-release for UVOT) calibration files\footnote{swxpc0to12s6\_20130101v014.rmf and swxwt0to2s6\_20131212v015.rmf for XRT; swugu0160\_20041120v105.arf and swugu0160wcal20041120v002.fits for the UVOT grism.}.
The standard grade selections (0--12 for Photon Counting mode -- PC; 0--2 for Windowed Timing mode -- WT) were chosen for XRT, with an annulus used for the extraction of source counts in order to avoid pile-up\footnote{When a source is super-soft, pile-up becomes obvious at lower count rates, so it is advisable to exclude more of the PSF than suggested for the harder sources discussed by Romano et al. (2006).} when the PC count rate was $\ga$0.3~count~s$^{-1}$ or when the WT rate was $\ga$30~count~s$^{-1}$. The annular exclusion radii ranged between 2 and 8 pixels for the PC data, while 2 core pixels were excluded from the WT data at the peak of the emission; the outer radius was set to 20 pixels (1 pixel = 2.36 arcsec). The spectra were binned to have a minimum of 1 count~bin$^{-1}$ to facilitate Cash statistic (Cash 1979) fitting within {\sc xspec} (Arnaud 1996).
Despite the X-ray spectra for V745~Sco not being in the low count regime, the C-statistic is still preferred since it provides less biased parameter estimates (e.g. Humphrey, Liu \& Buote 2009).
The observations during the main interval of the SSS emission (i.e. those datasets for which the spectra are fitted in Section~\ref{specanal}) were entirely in WT mode.

The UVOT grism spectra were extracted using the {\sc uvotpy} program (Kuin 2014) which implements the {\em Swift} UVOT grism calibration described in Kuin et al. (2015).  The wavelengths of the spectra 
were shifted by a few \AA\, to correct for the errors in 
anchor position which affect the wavelength scale origin.  Bad data were flagged by inspecting the 
grism images for zeroth order contamination.

The {\em Swift} X-ray and UV light-curves and X-ray hardness ratio are shown in Fig.~\ref{lc}, together with optical ($BVR_CI_C$) and IR ($JHK$) data obtained from SMARTS and AAVSO.

\subsection{SMARTS}

Optical and IR data were obtained using the ANDICAM (A Novel Double-Imaging CAMera) dual-channel imager on the SMARTS 1.3~m telescope and downloaded from the online Stony Brook/SMARTS Atlas of (mostly) Southern Novae (Walter et al. 2012). ANDICAM obtains simultaneous optical and IR data through the use of a dichroic filter (more details given in Walter et al. 2012).  Bias subtraction and flat-fielding are performed by the SMARTS pipeline before the data are distributed. 
Data in all seven filters were collected on most nights from 1.7~days after the outburst until around T+42~day, and then less frequently. The magnitudes up until day~54 are shown in the bottom panel of Fig.~\ref{lc}.

Fig.~\ref{lc} also includes early time (up to day 1.1) $BVRI$ photometry obtained from the AAVSO, to supplement the SMARTS data.

\section{X-ray spectral evolution}
\label{specanal}

\subsection{Pre-SSS}

\begin{figure*}
\begin{center}
\includegraphics[clip, angle=-90, width=14cm]{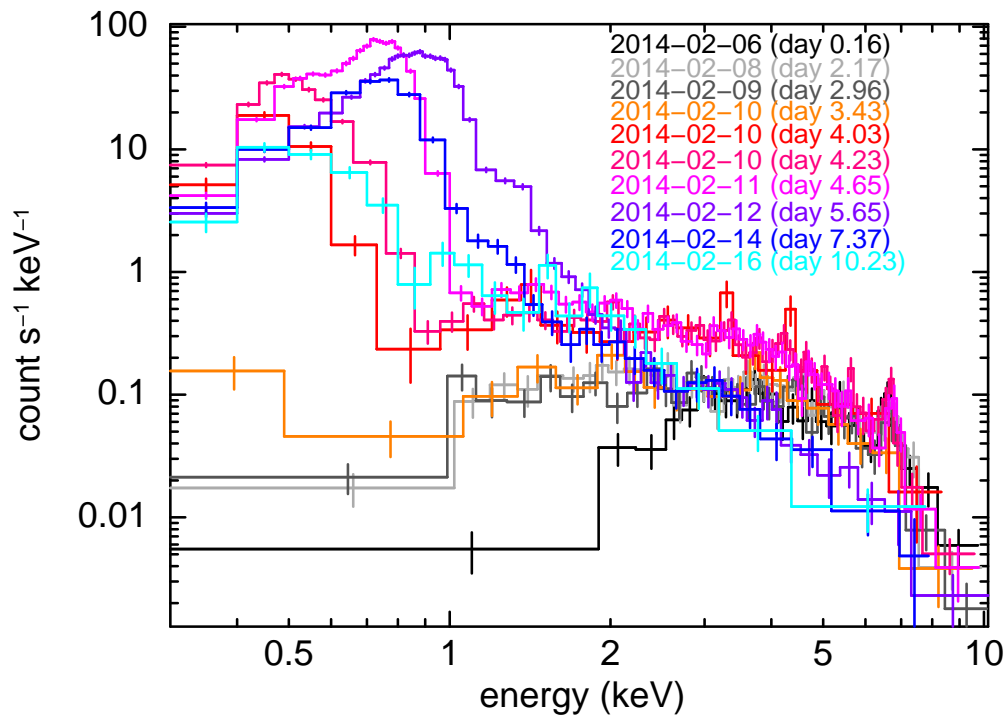}
\caption{A sample of X-ray spectra obtained during the SSS phase of V745~Sco. The legend to the right of the plot gives the date each spectrum was taken. The energy bins for each spectrum have been joined together by a line to help guide the eye.}
\label{specsample}
\end{center}
\end{figure*}

The X-ray spectral evolution of V745~Sco is striking, as shown in Fig.~\ref{specsample}. The first three spectra in the figure (days 0.16--2.96, plotted in greyscale) show the X-rays prior to the detection of any super-soft emission. The X-ray spectra at this time are hard, with little emission below 1--2~keV, due in part to a large amount of absorption.

\begin{figure}
\begin{center}
\includegraphics[clip, angle=-90, width=8.5cm]{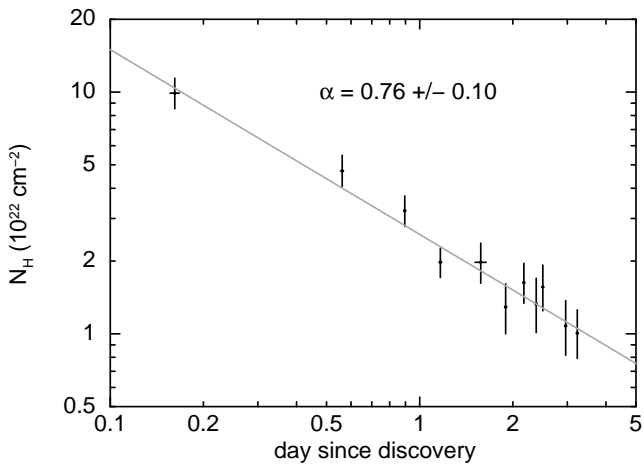}
\caption{Temporal behaviour of the nova ejecta/RG wind absorbing column. The absorption declines following a power law, N$_{\rm H}$~$\propto$~t$^{-0.76\pm0.10}$.}
\label{nh}
\end{center}
\end{figure}

The early spectra (before day 3.4) can be well-fitted with a single temperature optically thin component and two absorption parameters, one to model the expected interstellar N$_{\rm H}$ and the other to parametrize the declining column as the nova ejecta thin and the shock emerges from the secondary star wind (see Table~\ref{early}). $E(B-V)$ is given as 1.0~$\pm$~0.2 in Schaefer (2010), corresponding to N$_{\rm H}$~$\sim$~(5.6~$\pm$~1.1)~$\times$~10$^{21}$~cm$^{-2}$ [Predehl \& Schmitt (1995), assuming the standard R$_{\rm V}$~=~3.1 for the Milky Way Galaxy]; this is used as the fixed interstellar absorption column. The iron abundance in the collisionally ionized plasma ({\sc vapec} model in {\sc xspec}; Smith et al. 2001) was set to be 0.51 Solar, as found by Orio et al. (2015) from fitting {\em NuSTAR} data; all other abundances were left at Solar. 
Fig.~\ref{nh} demonstrates the decline in N$_{\rm H}$ from the optically thick nova ejecta expanding into the shocked RG wind before the super-soft emission becomes visible. The variable absorption can be modelled with a power-law decline of N$_{\rm H}$~$\propto$~t$^{-0.76\pm0.10}$, similar to, if somewhat steeper than, the findings for RS~Oph by Bode et al. (2006), where it was concluded that a decelerating shock was traversing a RG wind with N$_{\rm H}$~$\propto$~r$^{-2}$ (where r is the radial distance from the site of the explosion).
While it is geometrically possible that there could be additional absorption reaching down to the radiating surface of the WD, there is no evidence for such a column in these data. 

\begin{table}

\caption{Fits to the X-ray spectra before the SSS emission became visible. In each case the model consists of an optically thin emission component absorbed by two columns: a fixed interstellar value of 5.6~$\times$~10$^{21}$~cm$^{-2}$ and the variable N$_{\rm H}$ listed in the table. The iron abundance of the {\sc vapec} component was set to be 0.51 Solar as described in the text.}

\begin{center}
\begin{tabular}{llll}
\hline
Day & {\sc vapec} kT  & N$_{\rm H}$  & C-stat/dof\\
    &  (keV) & (10$^{22}$ cm$^{-2}$)\\
\hline
0.16 & $>$46 & 9.9$^{+1.0}_{-0.8}$ & 230/311\\
0.56 & $>$53 & 4.6$^{+0.5}_{-0.4}$ & 297/315\\ 
0.90 & $>$55 & 3.0~$\pm$~0.3 & 339/353\\
1.17 & $>$61 & 2.0~$\pm$~0.2 & 475/510\\
1.57 & $>$43 & 2.0~$\pm$~0.2 & 256/332\\
1.90 & $>$61 & 1.4~$\pm$~0.2 & 326/373\\
2.17 & $>$60 & 1.7~$\pm$~0.2 & 382/406\\
2.38 & $>$52 & 1.4~$\pm$~0.2 & 281/297\\
2.50 & $>$55 & 1.6~$\pm$~0.2 & 297/346\\
2.96 & $>$60 & 1.1~$\pm$~0.2 & 346/359\\
3.23 & $>$62 & 1.0$^{+0.2}_{-0.1}$ & 406/447\\

\hline
\end{tabular}

\label{early}
\end{center}
\end{table}

It should be noted that the measured N$_{\rm H}$ column is dependent on the abundances used. While a value of Fe/Fe$_{\odot}$~=~0.51 has been assumed for the fits, based on the {\em NuSTAR} spectral results (Orio et al. 2015), these earlier {\em Swift} data are actually better fitted with an enhanced iron abundance (with Fe/Fe$_{\odot}$~$\sim$~6). Such a dramatic drop in iron abundance is not expected to occur physically, and the apparent difference is likely a symptom of a more complicated situation -- for example, the system may be in non-equilibrium ionization, or there could be a distribution of temperatures contributing to the shock emission, leading to a thermal line blend. Reflection from the WD surface could also enhance the iron emission, leading to a neutral Fe fluorescence line of $\sim$~100~eV equivalent width for a column of N$_{\rm H}$~$\sim$~10$^{23}$~cm$^{-2}$ (Makishima 1986). Similar strong iron emission was also seen in the early, absorbed RS~Oph spectra (Bode et al. 2006). Further investigation into these early shock-emission spectra is beyond the scope of this paper. 

We also tried varying the oxygen abundance of the absorption. Decreasing (increasing) this to 0.5 (1.5) Solar increases (decreases) the column by 10--20~per~cent. Within the uncertainties, however, the N$_{\rm H}$ values remain consistent over this range of abundances.

Following Bode et al. (2006), Banerjee et al. (2014b) calculate that the shock temperature at 1.3~days after the nova explosion should be $>$10$^{8}$~K (i.e. $>$8.6 keV). {\em Swift} observed V745~Sco on days 1.2 and 1.6; fitting these spectra as detailed above provides shock temperatures of $>$40~keV, which is not inconsistent with the Banerjee prediction. 
The shock emission continues through the SSS onset and evolution (although swamped by the soft X-ray photons)
and is seen to be responsible for the 2--10~keV light curve shape. 
As the shock velocity declines, so, too, should the temperature of the shock emission; the {\em Swift} data only provide lower limits on the shock temperature {(see Table~\ref{early}), however, so this cannot be confirmed.

\subsection{SSS phase}

By day 3.4 (orange spectrum in Fig.~\ref{specsample}), soft X-ray emission can be seen starting to rise and, about fourteen hours later, as shown by the red spectrum, the soft photons were dominant. The shape and temperature of the soft emission changed dramatically during the SSS phase, as was also clearly seen in RS~Oph (Osborne et al. 2011), with the peak of the soft emission in V745~Sco moving to higher energies (corresponding to hotter temperatures) between days 4.0 and 5.6. It is also evident that the harder, shock emission appears to decrease in strength after day five or so.

X-ray spectra were extracted and fitted for each snapshot of data during the interval for which SSS emission was detected (days 4.0--12.5). Snapshots are typically between 500~s and 1.5~ks in duration. 

Although different TMAP\footnote{T{\" u}bingen NLTE Model Atmosphere Package: http://astro.uni-tuebingen.de/\raisebox{.2em}{\tiny$\sim$}rauch/TMAF/flux\_HHeCNONeMgSiS\_gen.html} (plane-parellel, static, non local thermal equilibrium) atmosphere models were fitted to the data, many of the spectra required a higher temperature than is covered by the available grids (the upper limit is 1.05~$\times$~10$^{6}$~K~=~90.5~eV). Therefore, despite the known problems with parametrizing the SSS emission with blackbodies (specifically: underestimation of the temperature and overestimation of the luminosities; see, e.g., Krautter et al. 1996), they still provided better fits than the atmosphere models, as shown in Fig.~\ref{atmos-bb}. Henze et al. (2011) showed for their sample of M31 novae that there was a strong correlation between the temperatures estimated using blackbody (BB) fits and those from model atmospheres, concluding that BBs can be used to characterise the SSS temperature changes. Osborne et al. (2011) actually found that the bolometric luminosities estimated for the peak X-ray `plateau' of RS~Oph were surprisingly the same for both BB and atmosphere fits, though BBs did provide order of magnitude higher luminosity values compared to the atmosphere grids during the high-amplitude variability phase. For the luminosity estimates in this paper, the distance to V745~Sco was taken to be 7.8~kpc from Schaefer (2010). 

\begin{figure}
\begin{center}
\includegraphics[clip, angle=-90, width=8.5cm]{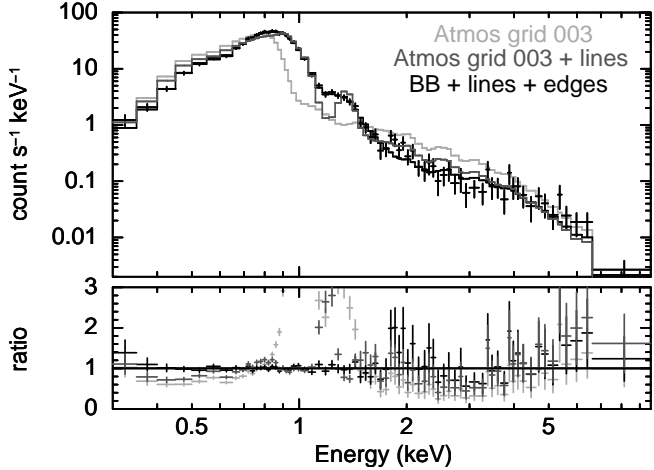}
\caption{An example spectrum from close to the peak of the SSS phase (day 6.7) fitted with both an atmosphere grid and a blackbody and, in both cases, an optically thin component. The atmosphere model provides a much worse fit overall compared to the blackbody, with or without the five emission lines (see text for details).}
\label{atmos-bb}
\end{center}
\end{figure}

\begin{table}

\caption{Comparison of fits to the day 6.7 spectrum (see Fig.~\ref{atmos-bb}). In each case, an optically thin {\sc vapec} component is also included. The BB+5lines+2edge provides the best result. Atmosphere grids 003 and 011 have differing abundances, as described in the text. Adding edges to the atmosphere model does not improve those fits. For the atmosphere models, the {\sc vapec} temperature was fixed at 2~keV, to model the higher energy emission.}

\begin{center}
\begin{tabular}{ll}
\hline
Model & C-stat/dof\\
\hline
BB & 867/317 = 2.74\\
BB+5lines+2edges & 266/310 = 0.86\\
Atmos 003 & 4302/318 = 13.53\\
Atmos 003+5lines & 671/313 = 2.14\\
Atmos 011 & 4596/318 = 14.45 \\
Atmos 011+5lines & 756/313 = 2.42\\

\hline
\end{tabular}

\label{fitcomp}
\end{center}
\end{table}


In each case the complete spectral model includes two absorbing columns: the fixed interstellar N$_{\rm H}$ and a variable value for the thinning nova ejecta, estimated by extrapolating the fit in Fig.~\ref{nh} to the relevant day for each observation.
In addition to the BB component used to parametrize the soft emission, an optically-thin {\sc vapec} component to account for the shock emission above $\sim$~2~keV, five Gaussian emission lines and two edges -- a cold, neutral oxygen edge at 0.54 keV and an H-like oxygen edge at 0.87~keV -- were included. The line energies were chosen to cover the strongest noticeable emission and were set at 0.57~keV, 0.65~keV (He-like and H-like O), 0.92~keV, 1.02~keV (He-like and H-like Ne) and 1.35~keV (He-like Mg). The addition of these lines and edges much improved the fit compared to a simple BB (Table~\ref{fitcomp}), resulting in a significant reduction in the C-statistic for each of the extra components for at least some of the spectra. 
 This model is clearly empirical and may not uniquely provide the best fit (though note that {\em Chandra} X-ray grating spectra detected some of the same emission lines, among others; Drake et al. 2014); however, the components chosen are not unreasonable and no likely alternative empirical model was found to provide better results. Despite both neutral and H-like O edges generally being statistically significant, He-like O (0.74~keV) was not. For the day 6.7 spectrum shown in Fig.~\ref{atmos-bb}, the 90\% upper limit on the optical depth was 0.09. At the peak of the SSS phase, $<$~4~per~cent of the 0.3--2~keV flux was contributed by the optically-thin plasma.

Including these same emission lines within the atmosphere grid model vastly improves these fits (compared to the atmosphere grid alone; see also Orio et al. 2015). This suggests that some of the abundances in the atmosphere grids are not appropriate for this nova. Table~\ref{fitcomp} lists the fit comparisons for the data from day 6.7 (see also Fig.~\ref{atmos-bb}) for reference. The main differences between atmosphere grids 003 and 011 are the abundances of nitrogen (a factor of $\sim$~40 higher in grid 003) and carbon (factor of $\sim$14 lower abundance in grid 003 compared to 011). The abundances are fixed and cannot be varied within any single atmosphere grid.

The evolution of the parameters from the BB fitting is shown in Fig.~\ref{specfits}. Prior to day 4.0 and after day 12.5, the inclusion of a soft component (be it a BB or an atmosphere model) does not improve the overall fit significantly. The final two spectra fitted (those after day 11) were each made up of three observations merged together, to obtain better signal-to-noise as the super-soft emission faded. We note that the bolometric correction (calculated as the ratio of the bolometric and in-band X-ray luminosities) for the first three spectral fits is extremely large (3.1~$\times$~10$^{4}$ for the first spectrum; the luminosity from this bin has been excluded from the plot), and should be regarded with caution.

\begin{figure}
\begin{center}
\includegraphics[clip, angle=-90, width=8.5cm]{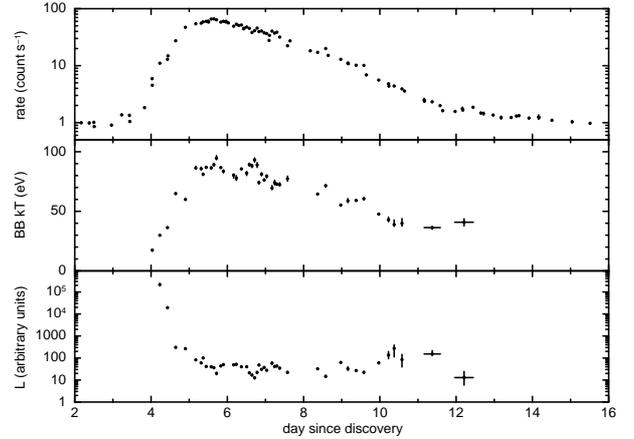}
\caption{Fits to the super-soft X-ray spectra of V745 Sco, consisting of a BB, an optically thin {\sc vapec} component, two oxygen edges (0.54 and 0.87 keV) and five Gaussian emission lines (fixed at 0.57, 0.65, 0.92, 1.02 and 1.35~keV), with variable absorption determined from a model to the declining ejecta and wind. The top panel shows the 0.3--10~keV X-ray light-curve; the second and third panels show the fitted BB temperature and derived bolometric luminosity (in arbitrary units). The luminosity for the first spectrum ($\sim$~2~$\times$10$^9$) has not been plotted, in order to compress the ordinate scale.}
\label{specfits}
\end{center}
\end{figure}


Fig.~\ref{lines} shows the temporal variation of the strengths of the emission lines (in terms of equivalent width) and edges. The 0.65 keV (H-like O) and 1.35 keV (He-like Mg) lines are significant throughout most of the SSS phase, while the Ne lines at 0.92 keV and 1.02 keV are strongest at the time of peak X-ray emission. The 0.57 keV (He-like O) line is only intermittently required.

The ionized O{\sc viii} edge at 0.87 keV is more significant in all the spectral fits (though decreasing in optical depth as the X-rays fade) compared to the O{\sc i} edge at 0.54 keV (only required during the peak emission interval).
Fig.~\ref{cont} demonstrates that the high values measured for the 0.92~keV Ne line and the 0.87~keV O edge on day 5.2 are strongly preferred; there is no good fit where the strength of the line and/or the edge can be close to zero. The lower plot shows the strong residuals remaining if the O absorption edges and Ne line are excluded from the model. Comparison with non-equilibrium ionization collisional plasma models shows that such a strong emission line would not result from time-dependent ionization effects. Similarly, the strengths of the 0.57 keV emission line and 0.54 keV absorption edge are strongly correlated, though both parameters are at times significant improvements to the overall fit.

\begin{figure*}
\begin{center}
\includegraphics[clip, width=12cm]{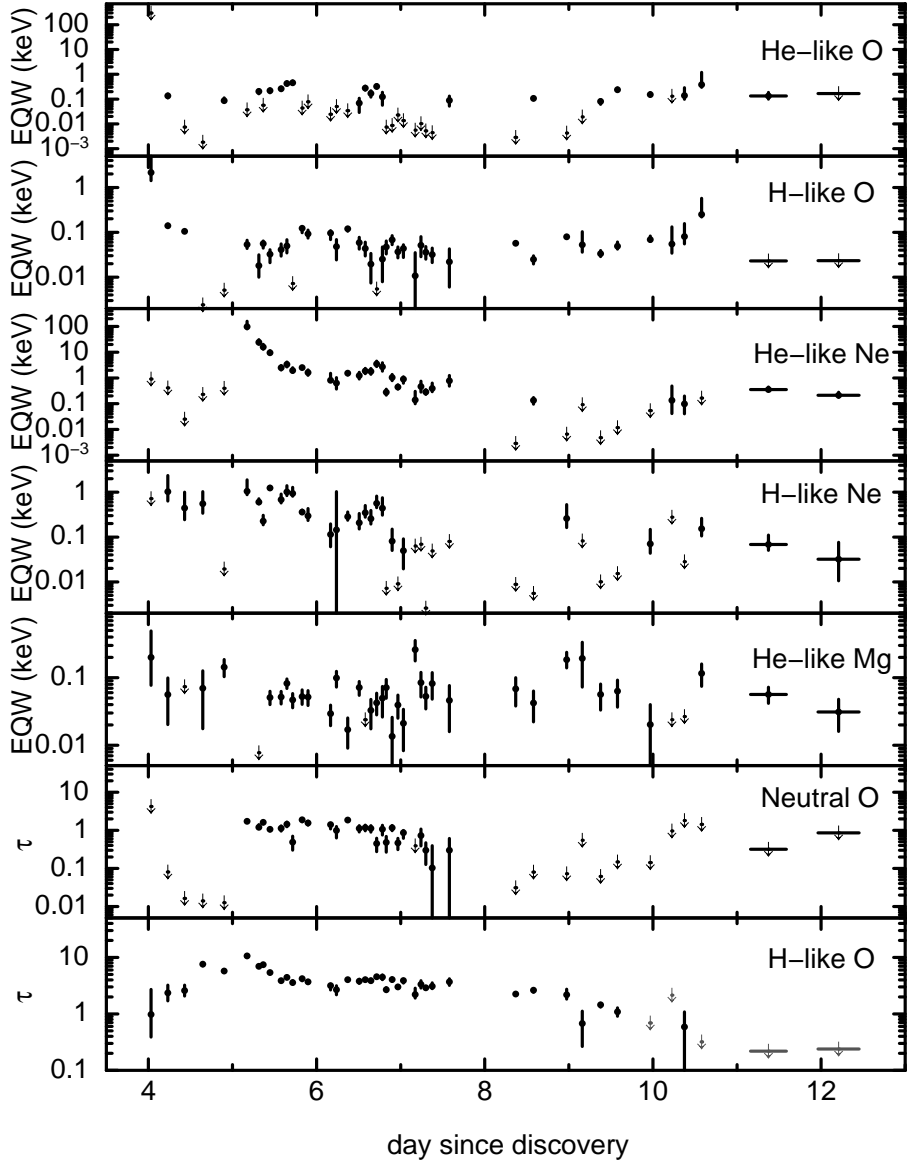}
\caption{The variation of the line equivalent widths and edge optical depths with time. Lines from He-like and H-like O are at energies of 0.57~keV and 0.65~keV, from He-like and H-like Ne at 0.92~keV and 1.02~keV and He-like Mg at 1.35~keV. The neutral and H-like O edges are at 0.54~keV and 0.87~keV respectively.}
\label{lines}
\end{center}
\end{figure*}

\begin{figure}
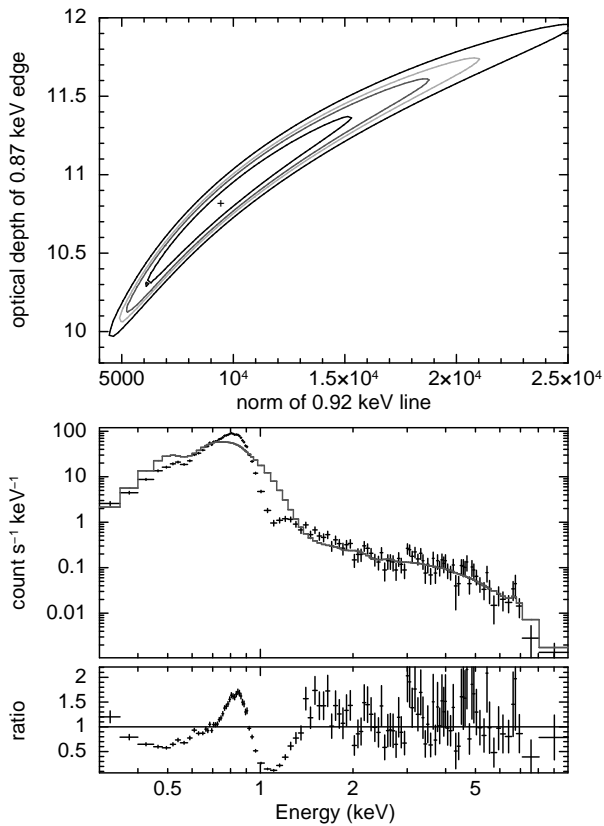

\begin{center}
\includegraphics[clip, angle=-90, width=8cm]{fig7.ps}
\includegraphics[clip, angle=-90, width=8cm]{fig7a.ps}
\caption{Top: A contour plot demonstrating that the spectrum from day 5.2 is significantly better fitted with both a very strong emission line at 0.92~keV and a deep absorption edge at 0.87 keV. The contours plotted are for the (inner to outer) 68, 90, 95 and 99~per~cent confidence intervals. For comparison with Fig.~\ref{lines}, the conversion between normalisation and equivalent width (in keV) is approximately 0.012: that is, the lower bound of a normalisation of 5000 equates to $\sim$~60~keV. Bottom: A fit to the day 5.2 spectrum  {\em excluding} the two absorption edges (0.54 and 0.87~keV) and the strong 0.92~keV emission line; the resulting fit is very poor and strong residuals are apparent.}
\label{cont}
\end{center}
\end{figure}

The temperature of the soft emission is seen to follow the shape of the X-ray light-curve: i.e., as the X-ray flux gets brighter, the BB temperature increases, and as the soft X-ray emission decreases, so, too, does the temperature. Fig.~\ref{kT-CR} demonstrates the relationship between the X-ray count rate and the BB temperature. There is a hysteresis effect, whereby a given count rate corresponds to a higher temperature at later times, as the source starts to fade and cool. This can be explained by the larger photospheric radii at the start of the SSS phase; by the later stages of the evolution, the effective radius has decreased and we can see deeper in, to the hotter layers.

\begin{figure}
\begin{center}
\includegraphics[clip, width=9cm]{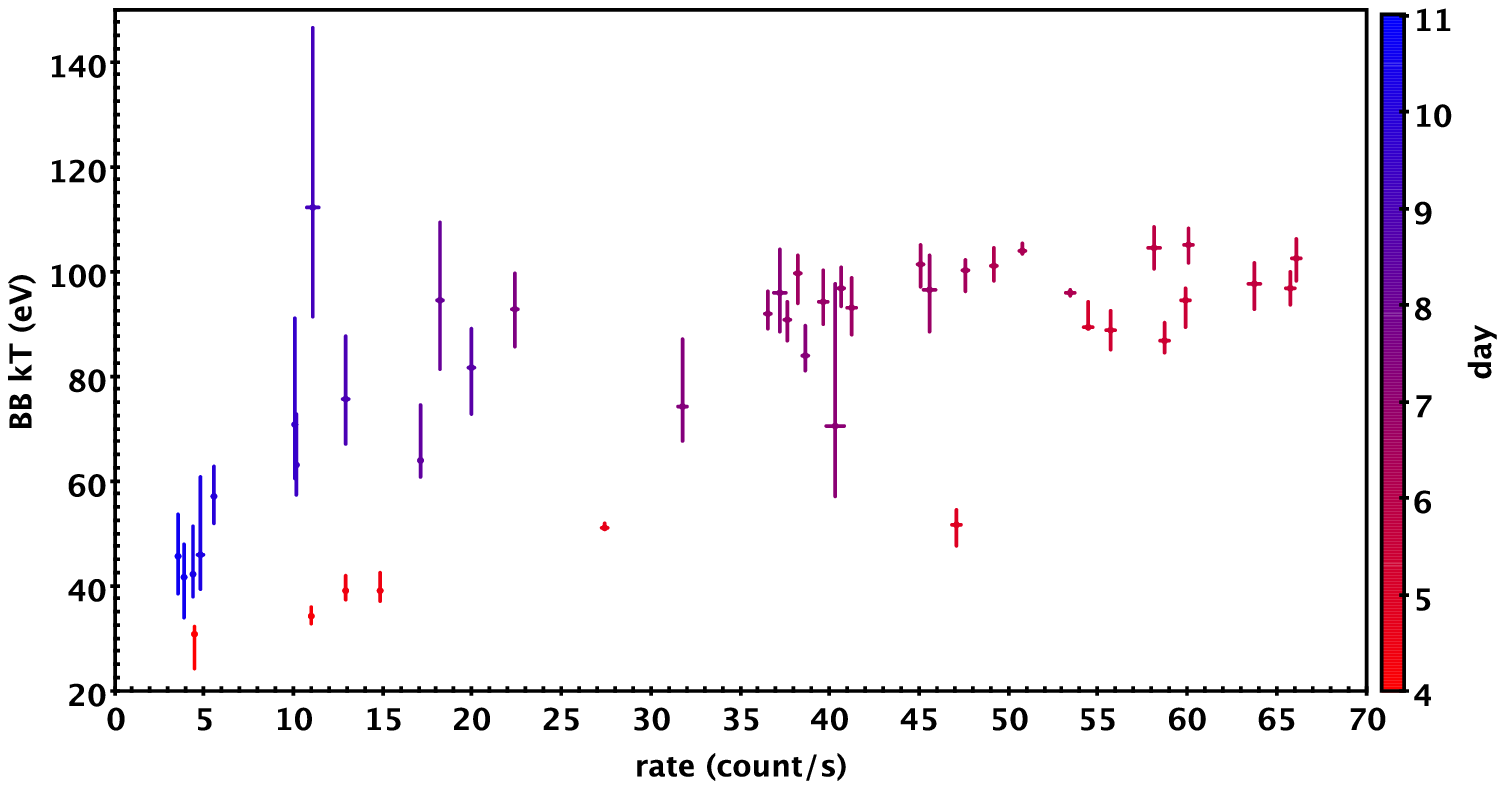}
\caption{The variation in BB temperature with X-ray count rate. The colour scale shows the passage of time, from the start (red) through to the end (blue) of the SSS phase.}
\label{kT-CR}
\end{center}
\end{figure}

Beardmore et al. (2014) reported a probable quasi-periodic oscillation in the {\em Swift} X-ray data. However, a more detailed analysis (Beardmore et al. in prep) no longer finds this to be a significant detection.

\section{UV spectral evolution}

Fig.~\ref{grism} shows a sample of the UVOT grism spectra, demonstrating the
evolution over time.
The spectra show mainly emission lines corresponding to lower ionization transitions, with blended fluorescent lines from N{\sc iii} and C{\sc iii} being mostly responsible 
for the feature at 4650~\AA. 
Shortward of 4000~\AA, the Balmer continuum is in emission, and varies with time, both in intensity and shape: this is particularly clear between days 4.5 and 5.5 after outburst (yellow and navy blue spectra in Fig.~\ref{grism}). This coincides with the time of the appearance and rapid brightening of the soft X-ray component, and is most likely due to recombination driven by the hot radiation 
field from the WD below. 
The spectra in Fig.~\ref{grism} were smoothed with a 3-point boxcar method, equivalent to around 10\AA\, in the UV. 


More detailed analysis of these grism data, together with SMARTS/Chiron spectra, will be presented in Kuin et al. (in prep).

\begin{figure*}
\begin{center}
\includegraphics[width=15cm]{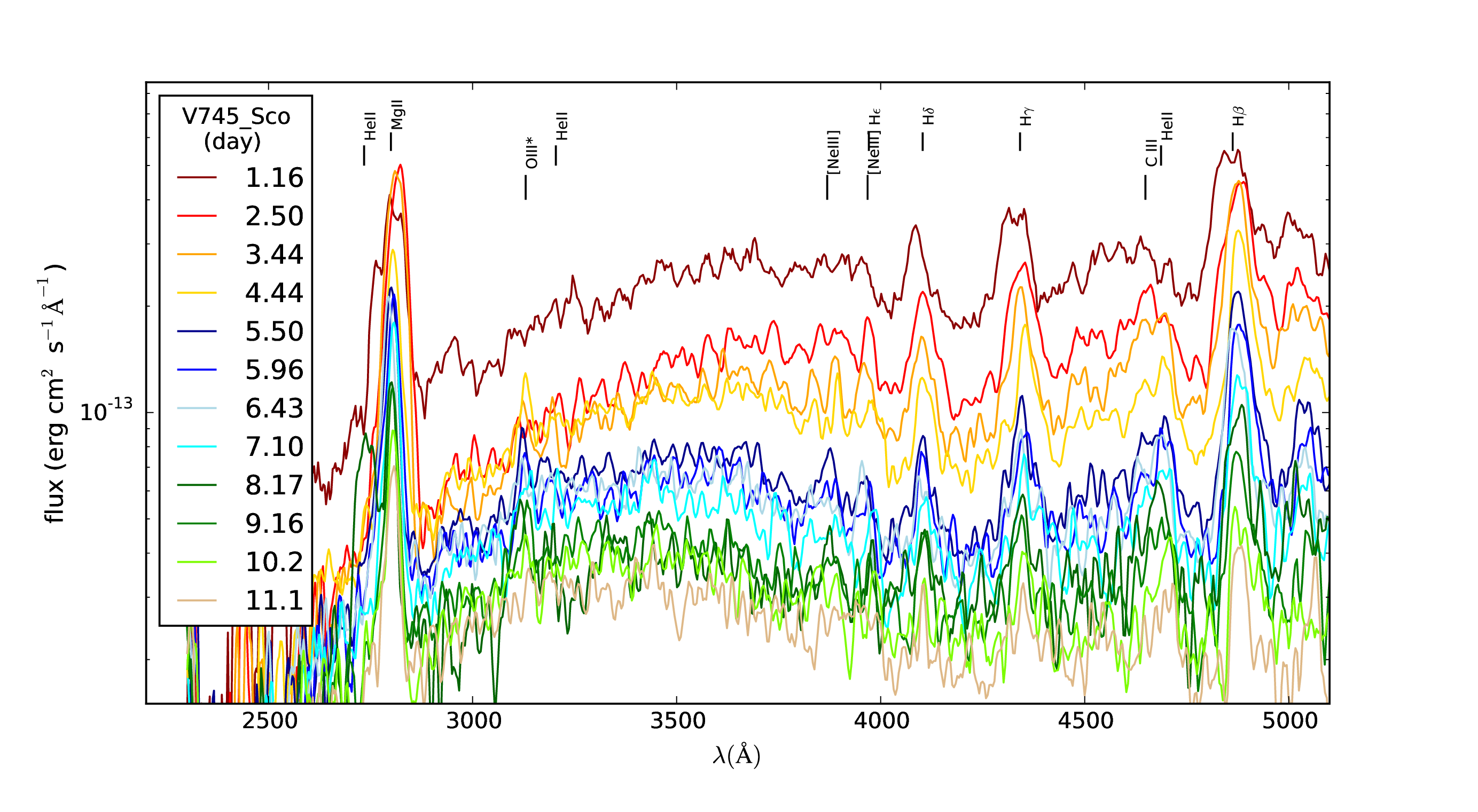}
\caption{{\em Swift} UV grism spectra plotted in log space, showing the evolution over time. Strong emission lines are labelled.}
\label{grism}
\end{center}
\end{figure*}

\section{Light-curve evolution}
\label{decay}

In contrast to the early increase in X-ray flux, the UV and optical emission faded continuously, though with changes in the rate of decay during the evolution of the nova. Following Page et al. (2013), we parametrize the evolution of the magnitudes over extended intervals as being proportional to log(time), i.e. $f~\propto~(t/\rm {1~day})^{-\alpha}$.  When data from all three UV filters were available (before day 70), they each provided similar slopes, so the mean values are reported here.
The very earliest data ($\sim$~days 0.4--6) can be approximated with $\alpha$~$\sim$~0.7. The decay then steepens, to $\alpha$~$\sim$~3 until about day 19, after which time the slope is more gradual, with $\alpha$~$\sim$~1.4 until around day 70. Beyond this time, only the $uvw1$ filter provided detections, but these data indicate another steepening, to $\alpha$~$\sim$~2.5. As a guide line, the statistical errors on the magnitudes are below $\sim$~1~per~cent.

The optical curves appear to show more gradual changes in slope. After around day 4--5, the $BVR_CI_C$ curves steepened from $\alpha$~$\sim$~1 to $\sim$~2, though this is somewhat flatter overall ($\sim$~0.6 changing to 1.2) in the $R$-band. A further steepening to $\alpha$~$\sim$~4--5 occurs about day 10, after which the decay steadily flattens, ending with slopes of between $\alpha$~$\sim$~1.2 ($B$-band) and $\sim$~0.17 ($I$-band) between days 46 and 54. The IR data (see also Banerjee et al. 2014b) reveal a `plateau' about days 4--9, likely caused by reprocessing of the bright, SSS X-rays.
The flattening of the IR seen in Fig.~\ref{lc} after around day 12 results from the increasing dominance of the RG emission in the V745~Sco system as the light from the nova outburst fades away.

The time at which the UV slope changes from flat to steep (around day 6) is when the SSS X-ray  emission starts to fade from its maximum. The flattening at around day 19 is also evident in the X-ray decay (change in slope from $\alpha$~=~2.3~$\pm$~0.1 between days 12--19, to 1.18~$\pm$~0.03 at later times when considering the soft, 0.3--2~keV band). After around day 70, the X-ray slope also steepens again, to $\alpha$~$\sim$~2.6~$\pm$~0.2. 
It is worth noting that the decay slopes after day 19 are comparable in the X-ray and UV bands. This suggests that, despite the X-ray and UV light-curves being very different in early shape, there is a link between the emission regions. A similar relationship between the X-ray and UV decay curves was seen in V2491~Cyg (Page et al. 2010).

\section{Comparison with other novae}
\label{comp}

\subsection{V2491 Cyg}

The shape of the V745~Sco X-ray light-curve (Fig.~\ref{lc}) is strongly reminiscent of that found for V2491~Cyg (Page et al. 2010). Both novae show a steep rise to peak count rate (steeper than seen for any other nova followed by {\em Swift}; Schwarz et al. 2011), followed by an almost immediate rapid decline which then slows after fewer than 10 days. In both cases, the interval during which the X-ray emission stays at the peak is short, only a day or so. 
Other novae monitored in detail by {\em Swift} have tended to show a `plateau' for many days to weeks at the peak X-ray count rate (e.g. RS~Oph -- Osborne et al. 2011; KT~Eri -- Schwarz et al. 2011; T~Pyx -- Tofflemire et al. 2013, Chomiuk et al. 2014; Nova LMC 2012 -- Schwarz et al. 2015; Nova LMC 2009a -- Bode et al. in prep.), though sometimes modulated by the system's orbital period (e.g. HV Cet -- Beardmore et al. 2012; V959~Mon -- Page et al. 2013).

In Fig.~\ref{comp} the data for V745~Sco and V2491~Cyg are overplotted with no scaling in the count rate direction, but with a shift of $-$35.6~days applied for the V2491~Cyg dataset to align the X-ray peaks in time. The pattern followed by the count-rate light-curves is clearly very similar, while the hardness ratios (bottom panel) are divergent; this, however, is likely to be related, at least in part, to the differing secondary star wind environments. V2491~Cyg shows a `flare' around T+2 days (shifted time scale; 37.6~days in the actual observed time frame) in the light-curve plot, which may have been a brief interval of the variability sometimes seen at the start of the SSS phase (see Section~\ref{disc}). We note that this `flare' was not mentioned in Page et al. (2010) because of the different time binning used in that work. It was, however, shown in the subsequent paper by Ness et al. (2011).

The BB fits to the V2491~Cyg spectra presented by Page et al. (2010) peak at a slightly lower temperature than in V745~Sco ($\sim$~80~eV). This soft component remained significant, even as the X-ray emission faded, with little evidence of cooling. A {\em Suzaku} observation of the source in quiescence, two years after the outburst, still revealed a BB with kT~=77$^{+7}_{-9}$~eV (Zemko, Mukai \& Orio 2015). This is in contrast to the V745~Sco data presented here, which clearly show a cooling of the BB at the end of the SSS phase. The general trend of the luminosity evolution is the same for both objects, starting high, then fading to an approximately constant value, with the luminosity stabilising around the time of peak SSS temperature. Assuming a distance of 10.5~kpc, Zemko et al. (2015) find a luminosity of 1.4~$\times$~10$^{35}$~erg~s$^{-1}$ for V2491~Cyg in their 2010 data, which is the same order of magnitude as seen towards the end of the {\em Swift} campaign in Page et al. (2010). Although the luminosities are uncertain given the nature of BB fits, the V745~Sco spectra correspond to a luminosity of around 1.6~$\times$~10$^{39-40}$~erg~s$^{-1}$ following the plateauing after day five or so, several orders of magnitude brighter than V2491~Cyg. 


Page et al. (2010) suggested that V2491~Cyg could be a recurrent nova, given its pre-outburst detection in X-rays (Ibarra et al. 2009), though most likely with a timescale in the range of $\sim$~100~yr. We show here that the steep rise in SSS X-ray emission and the subsequent rate of decline are very similar for the proposed recurrent nova V2491~Cyg and the known recurrent V745~Sco. The onset of the detectable SSS emission is thought to correspond to the reduction in ejecta column density (e.g. Krautter 2008) - a speedy switch-on implying there is little ejecta to clear, particularly true for V745~Sco. A prompt decline in the X-ray emission after peak suggests there was little material available for nuclear burning. Both of these situations - a small amount of ejecta and fuel - indicate a high WD mass ($\go 1.2 \msun$), as expected for recurrent novae.

\begin{figure}
\begin{center}
\includegraphics[clip, angle=-90, width=8cm]{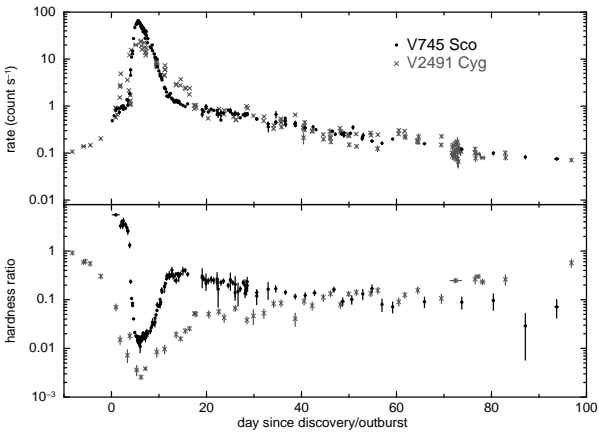}
\caption{A comparison of the 0.3--10~keV X-ray light-curves (top panel) and 2--10~keV/0.3--2~keV hardness ratios (bottom panel) of V745~Sco (black circles) and V2491~Cyg (grey crosses). The V2491~Cyg data have been shifted by $-$35.6 days, to align the X-ray peaks in time. No vertical scaling has been applied.}
\label{comp}
\end{center}
\end{figure}

\subsection{M31N 2008-12a}
\label{m31}

V745~Sco shows the earliest known detection of SSS emission after a nova outburst, at about four days. 
M31N 2008-12a (also known as M31N 2012-10a), a recurrent nova in M31 (Darnley et al. 2015), was found to evolve into a bright SSS six days after optical outburst in 2014 (Henze et al. 2014c,d, 2015), with the phase ending about 18.5 days after outburst (Henze et al. 2015e); a similar SSS turn-on rate was previously seen for the 2013 outburst (Henze et al. 2014b; Tang et al. 2014). Fig.~\ref{m31comp} compares the X-ray light-curves and BB fits to the SSS spectra for these two RNe. To aid the eye, the V745~Sco count rates (top panel) have been scaled down by a factor of 1800, while the BB temperatures (bottom panel) have been shifted up by 25 eV. The time axes have not been altered. This plot clearly demonstrates that, not only was the SSS detected earlier in V745~Sco (even allowing for the $\sim$~1~day uncertainty in the exact start time), the flux and SSS BB temperature also increased more abruptly than in M31N~2008-12a.  The rates of decline appear more similar, probably due to this stage of the evolution being simply dominated by the cooling of the WD (L~$\propto$~T$^4$). M31N~2008-12a did, however, first show a brief ($<$10~days) plateau at the highest count rate (which, as mentioned above, is more often than not seen in the X-ray light-curves of well-observed novae). 

We note that the BB fits performed by Henze et al. (2014b, 2015) did not include the absorption edges we use here. Removal of these edges leads the BB temperatures for V745 Sco being up to $\sim$~10~eV cooler than the values presented here.

Henze et al. (2010, 2011, 2014a) monitored novae in the central region of M31 between 2006 and 2012. They show that, for novae as fast as V745~Sco, the peak SSS effective temperature is typically significantly higher than we measure here. That is, we might expect V745~Sco to show temperatures around the same values as measured for M31N~2008-12a, rather than being $\sim$~25--35~eV cooler. Such a discrepancy between predicted and measured temperatures has also been identified in a recent M31 nova to be discussed by Henze et al. (in prep.). They propose that, by the time the opacity had diminished sufficiently for us to see deep enough into the photosphere, the nuclear burning had already ended and the SSS was cooling, hence the peak temperature was not observed. This could also explain the temperatures seen for V745~Sco. It would imply that the t$_{\rm off}$ was even earlier than the beginning of the visible decline in count rate ($\sim$~day~6) and therefore very close in time to t$_{\rm on}$ ($\sim$~day~4), the point at which the ejecta started to become transparent to X-rays. Thus, V745~Sco might be near to the so-called `unobservable region', or `invisibility zone', for novae, where the expanding ejecta lead to self-absorption of the SSS emission until the phase ends. Novae within this zone would never be observable as SSSs.

\begin{figure}
\begin{center}
\includegraphics[clip, angle=-90, width=8cm]{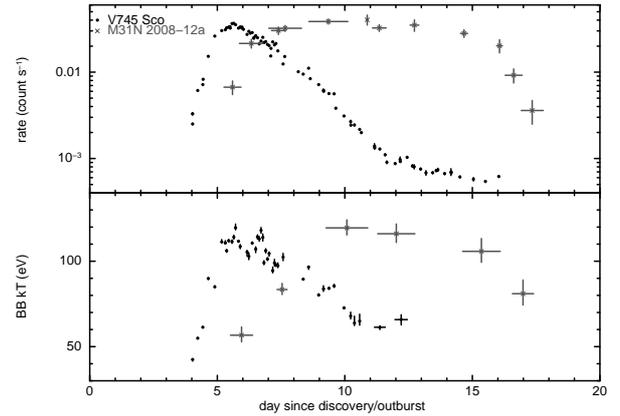}
\caption{A comparison of the 0.3-10~keV X-ray light-curves (top panel) and BB spectral fits for V745~Sco (black circles) and M31N 2008-12a (grey crosses; data from 2014 outburst). The V745~Sco count rates have been divided by 1800 and its BB temperatures shifted up by 25 eV in order to compare with the M31N 2008-12a curves.}
\label{m31comp}
\end{center}
\end{figure}

M31N 2008-12a has an extremely rapid recurrence time of $\sim$~1~yr, compared to $\sim$~25~yr for V745~Sco. Recently, Kato, Saio \& Hachisu (2015) presented a comprehensive theoretical light-curve model of the M31 nova which indicates a mass of 1.38~$\msun$ and a high accretion rate of 1.6~$\times$~10$^{-7}$~$\msun$~yr$^{-1}$. Using the luminosity-temperature models in Wolf et al. (2013; see also Sala \& Hernanz 2005), we estimate that the WD mass is $\sim$~1.3~$\msun$ for V745~Sco. However, Wolf et al. do not include WD masses in excess of this value. In addition, a higher peak SSS temperature as discussed above would suggest a larger WD mass from such models. By modelling the optical light-curve from the 1989 outburst, Hachisu \& Kato (2001) estimate the mass of V745~Sco to be 1.35~$\pm$~0.01~$\msun$, although they also derive a distance of 5.0~kpc, compared to the value of 7.8~kpc assumed here. In summary, 1.3~$\msun$ is a lower limit on the WD mass in V745~Sco.
Given that V745~Sco has both a shorter t$_{\rm on}$ (implying less mass ejected during the nova explosion) and a more rapid t$_{\rm off}$ (therefore less hydrogen burned) than M31N 2008-12a, the much slower recurrence time might suggest that the accretion rate is significantly less in the V745~Sco system, unless we have underestimated the WD mass for V745~Sco (as discussed above) or over-estimated for M31N 2008-12a. According to models presented by Yaron et al. (2015), a lower accretion rate should result in higher ejection velocities, such as those measured by Anupama et al. (2014) and Banerjee et al. (2014a,b) for V745~Sco. The accretion rate is considered further in $\S$~\ref{disc}.

We can use the similarities between V745~Sco and M31N 2008-12a to provide a rough, independent distance estimate. As in Fig.~\ref{m31comp}, scaling the V745~Sco count rates down by a factor of  $\sim$~1800 and shifting the temperatures up by a factor of $\sim$~1.25 (comparing the peak values of 95 and 120~eV for V745~Sco and the M31 nova respectively) provides a distance scaling factor of $\sim$~68. If we take the distance of M31 to be 780~kpc, this gives an approximate distance to V745~Sco of $\sim$~11~kpc (with a 95~per~cent uncertainty range of 8--14~kpc). This estimate does not account for differences in the absorbing column, yet is similar to the distance of 7.8~$\pm$~1.8~kpc given by Schaefer (2010).

\section{Discussion}
\label{disc}

The launch of {\em Swift} in 2004 has led to the detailed X-ray monitoring of the outbursts of an ever-increasing number of novae, revealing previously unexpected behaviour, particularly with regards to the super-soft emission: Schwarz et al. (2011) present a sample of super-soft novae observed by {\em Swift}; see also Page et al. (2010), Osborne et al. (2011), Beardmore et al. (2012) and Page et al. (2013) for specific examples. V745~Sco, followed from 3.7~hr after optical discovery through the onset and decay of the SSS phase, adds to this wealth of data.

It is noticeable that the rise in X-ray emission as the super-soft source began (as shown in the hardness ratio panel of Fig.~\ref{lc}) was monotonic, in contrast to the high-amplitude flux variability shown during this phase for some other novae well-monitored by {\em Swift} (e.g. V458 Vul -- Drake et al. 2008, Ness et al. 2009; KT~Eri -- Bode et al. 2010, Beardmore et al. 2010; RS~Oph -- Osborne et al. 2011; LMC 2009a -- Bode et al. in prep.). Given that V745~Sco is an RS-Oph-type RN, it is perhaps interesting that it does not show the same variability. A reasonable hypothesis for the early variability is the existence of dense, clumpy ejecta passing through the line of sight; in this case, the lack of such variability could reflect a smoother density distribution of the ejecta, or viewing angle effects. Alternatively, a short high-amplitude variability phase could have occurred before the ejecta became transparent to X-rays, as discussed in the previous section.

The V745~Sco X-ray spectral fits are significantly improved by the addition of strong oxygen, neon and magnesium emission lines [making it of the SSe subclass defined by Ness et al. (2013)]. Unfortunately, due to the speed of evolution of V745~Sco, no X-ray grating spectra were obtained during the SSS phase, meaning we cannot place firmer constraints on the ionization levels present at this time. The existence of these lines suggests that the WD in the V745 Sco system may be of the ONeMg type, and therefore unlikely to be a possible progenitor for a Type Ia supernova explosion (whether or not its mass is increasing with time). Future detailed analysis of the UVOT, SMARTS and {\em Chandra} grating spectra may provide further conclusions on this question.

As Fig.~\ref{lines} shows, the strength of some of these lines varied dramatically, with the strongest occurring during the interval of peak X-ray emission. Ness (2015) discusses grating spectra obtained during the time of the SSS phase in RS~Oph: there, as here, emission lines are clearly superimposed on the SSS continuum, and are stronger when the overall source is brighter, indicating that they are not just a continuation of the earlier shock emission, and photoexcitation effects may be involved.

The change in bolometric luminosity of the super-soft emission is shown in the bottom panel of Fig.~\ref{specfits}. The values are initially strongly super-Eddington (up to $>$10$^{5}$ L$_{\rm Edd}$) -- unphysically high -- which is likely an effect of using the simplified BB parametrization: underestimating the temperature of the SSS emission means that it will be more strongly affected by the absorbing column, requiring a large increase in effective radius (and, hence, estimated luminosity) to compensate.
For this reason, the luminosity in the figure has been plotted in arbitrary units. Spot checks of the few spectra for which the atmosphere grids provide a fit to the temperature suggest a peak luminosity of a few hundred L$_{\rm Edd}$, but we caution that these fits are statistically significantly worse than the BBs (see Table~\ref{fitcomp}).
The estimated luminosity drops sharply between days four and five, when the increase in BB temperature is seen. 
However, although frequently not directly observed, theory predicts (e.g. MacDonald, Fujimoto \& Truran 1985) that nuclear burning should continue at constant bolometric luminosity, depleting
the envelope mass and causing the envelope (and therefore the effective radius
of the photosphere) to contract and the temperature of the emission to increase. To investigate this, if we assume that the luminosity during the rise in temperature does actually remain constant at the Eddington value of $\sim$~3.2~$\times$~10$^{4}$(M/$\msun$)$\lsun$, and taking M~$\sim$~1.3~$\msun$ (as discussed above), then we find that the corresponding photospheric radius decreases from $\sim$~12--0.4~$\times$~10$^{9}$~cm during this interval. (This may be an underestimate of the peak luminosity, however, since novae can reach super-Eddington luminosities -- e.g., Schwarz et al. 2001.)
Such a small radius (i.e. 4~$\times$~10$^{8}$~cm) would imply a WD mass of $\sim$~1.2~$\msun$ from the Nauenberg (1972) mass-radius relation for non-accreting WDs. This derived value should be taken as an estimate of the minimum mass, since further contraction of the radius should occur.

Following Henze et al. (2010, 2011, 2014a), we can estimate the mass accretion rate in the system. Taking the expansion velocity to be 4000~km~s$^{-1}$ (Banerjee et al. 2014a), the mass of hydrogen ejected in the outburst, M$_{\rm ej}$, is $\sim$~1.6~$\times$~10$^{-7}$~$\msun$ (from equation 2 in Henze et al. 2014a, assuming, as they did, that the SSS turns on when N$_{\rm H}$ decreases to 10$^{21}$~cm$^{-2}$; in the case of V745~Sco, t$_{\rm on}$~$\sim$~4~days). The amount of hydrogen burned on the WD surface (equation 3 in Henze et al. 2014a) requires knowledge of the bolometric luminosity of the system. If we estimate this to be the Eddington luminosity for a 1.3~$\msun$ WD (=4.16~$\times$~10$^{4}$~$\lsun$), and take 6~days to be the turn-off time of the nuclear burning, then, assuming the hydrogen fraction of the burned material to be 0.5 and the energy released per gram of hydrogen processed to be 5.98~$\times$~10$^{18}$~erg~g$^{-1}$ (Henze et al. 2014a; Sala \& Hernanz 2005), we estimate M$_{\rm burn}$~=~1.4~$\times$~10$^{-8}$~$\msun$. Hence, the accreted mass, M$_{\rm acc}$~=~M$_{\rm ej}$+M$_{\rm burn}$, is approximately 1.8~$\times$~10$^{-7}$~$\msun$, which, over a 25 year recurrence period, equates to a mass accretion rate, {\. M}$_{\rm acc}$~$\sim$~7~$\times$~10$^{-9}$~$\msun$~yr$^{-1}$. This is, of course, an approximation, since the exact bolometric luminosity is unknown (may be higher than L$_{\rm Edd}$), as is the precise t$_{\rm off}$ value (may be shorter than 6~days; see $\S$\ref{m31}). However, it is about a factor of 20 lower than the accretion rate of 1.6~$\times$~10$^{-7}$~$\msun$~yr$^{-1}$ estimated by Kato et al. (2015) for M31N~2008-12a.

Another aspect to consider regarding the differences between V745~Sco and M31N~2008-12a is that the viewing angle of the two systems could be different. If, for example, V745~Sco is viewed pole-on, the ejecta should be able to expand freely, quickly becoming transparent to X-rays. Given the high accretion rate for M31N 2008-12a, there is likely to be a large circumbinary disc. If the system is oriented edge-on to our line of sight, the soft X-rays would have to pass through this extra material before becoming observable, taking $>$~3~days to do so from the 2014 observations (Darnley et al. 2015). This, together with the lower accretion rate estimate, could go some way to explaining why t$_{\rm on}$ is shorter in V745~Sco than M31N~2008-12. This could also be the reason why V2491~Cyg only became visible as a SSS around 35.6~days later than V745~Sco, despite the light-curve shape being very similar: Ribeiro et al. (2011) give an inclination angle of 80$^{+3}_{-12}$~deg for V2491~Cyg.

The $\sim$~25~yr recurrence time of V745~Sco is similar to that of RS~Oph (20~yr), though the SSS evolution occurs much more quickly in V745~Sco. In the case of RS~Oph, the first hint of the SSS was 26~days after outburst (compared with four days in V745~Sco), with the soft emission starting to fade after around day 60 (Osborne et al. 2011), rather than day six. Osborne et al. (2011) found atmosphere model temperatures peaking about 90~eV in RS~Oph, though BB fits (e.g. Page et al. 2008) showed lower peak temperatures of about 60~eV. The peak bolometric luminosity found for RS~Oph (from either atmosphere or BB fits; Osborne et al. 2011; Page et al. 2008) were around 1--2~L$_{\rm Edd}$, orders of magnitude lower than the values found from the BB fits to V745~Sco ($>$10$^{4}$~L$_{\rm Edd}$). RS~Oph is believed to have a WD mass close to the Chandresekhar limit (Osborne et al. 2011 and references therein), although this comparison suggests that the V745~Sco system may harbour an even more massive star.

Luna et al. (2014) analysed an {\em XMM-Newton} observation of V745~Sco in quiescence, finding a count rate of 3.6~$\times$~10$^{-4}$ count~s$^{-1}$ in EPIC-MOS, which corresponds to $\sim$~10$^{-4}$ count~s$^{-1}$ in {\em Swift}-XRT PC mode, more than an order of magnitude below the brightness of the final observation reported here. If the current rate of decay seen in the X-rays continues ($\S$~\ref{decay}), it will take until $\sim$~1000~days post-outburst before the source is back at the quiescent X-ray level.

V745~Sco was marginally detected by the {\em Fermi}-LAT (Cheung et al. 2014a). Of all the other LAT-detected novae, V745~Sco is most like V407~Cyg (Abdo et al. 2010; Shore et al. 2011), both being symbiotic systems with the WD embedded in a wind from its companion star. Tatischeff \& Hernanz (2007) predicted that particle acceleration would have occurred in the symbiotic recurrent nova RS Oph, implying TeV gamma-rays would be produced; however its last outburst
occurred before {\em Fermi} was launched. Abdo et al. (2010) proposed that the $\gamma$-rays detected from V407~Cyg were formed by Fermi-accelerated particles caused by the nova ejecta shock against the RG wind, with either $\pi^0$ decay or electron inverse-Compton scattering of the nova light forming the spectrum. The latter was favoured in spectral fits by Martin \& Dubus (2013), who also showed that the
observed $\gamma$-ray light curve requires the presence of a mass accumulation close to the WD, likely in
the equatorial plane. Metzger et al. (2015), building on the Chomiuk et al. (2014) model of shocks forming at the
interface of a fast spherical outflow with a slower outflow in the orbital plane (or perhaps at ejecta clumps),
favoured a hadronic origin. These authors noted that the associated X-ray emission is apparently absorbed below
detectability at energies less than 10~keV, and that hard X-ray observations may constrain the shocks. Orio et al.
(2015) found no X-ray emission between $\sim$~20--79 keV in {\em NuSTAR} observations of V745~Sco which, together with an
inability to fit the {\em NuSTAR} spectrum with a power-law, allowed them to rule out Comptonised $\gamma$-rays forming hard X-rays ten days after outburst. A stronger test of the Metzger et al. model requires hard X-ray observations
to be made earlier, at the time of the $\gamma$-ray peak.

\section{Summary}

Nova V745~Sco (2014) is the fastest evolving nova in terms of the SSS phase to date. As such, it provides a useful environment in which to study all the different stages of X-ray emission.

V745~Sco was marginally detected at $>$100~MeV, making it the sixth $\gamma$-ray nova detected by the {\em Fermi}-LAT.
 The nova showed the earliest detected switch-on of super-soft emission, at $\sim$~4~days after outburst (although the outburst could have occurred up to 24~hr earlier), followed rapidly by the turn-off around two days later (or possibly even sooner, depending on whether or not the time of the peak X-ray temperature was observed). This very early start to the SSS phase implies that any ejecta from the nova explosion were both low in mass and high in velocity, suggesting a high mass WD in this recurrent nova system.

During the brief interval before the SSS phase, the X-ray spectra are well fitted by shock emission absorbed both by the interstellar column and a declining column due to the expanding nova ejecta and RG wind. During the visible SSS phase, the soft thermal component increases in temperature to a peak of around 95~eV, as expected from a high white dwarf mass ($>$1.3~$\msun$). Throughout this phase, oxygen, neon and magnesium emission lines, and oxygen absorption edges, superimposed on a blackbody-like continuum, show significant evolution, with emission from helium-like neon being particularly strong at the time of peak X-rays, suggesting that the WD in the V745~Sco system may be of the ONeMg type.

The UV grism spectra also show considerable evolution, with changes in the continuum level occurring when the soft X-ray component first became visible. Comparison of the change in decay slopes of both the UV and X-ray light-curves indicates there is a link between the separate emission regions, despite the distinctly different shapes of the light-curves early on. 

V745~Sco shows a very similar X-ray light-curve shape to that of V2491~Cyg, though with the super-soft emission starting more than 35~days earlier in V745~Sco. The light-curve and spectral fits are also similar to those obtained for the shortest known inter-outburst-period recurrent nova M31N~2008-12a, although V745~Sco evolves more quickly throughout a single outburst. Estimates suggest that the mass accretion rate in V745~Sco could be $\sim$~20$\times$ lower than in M31N~2008-12a, which might explain why the SSS emission becomes visible more rapidly than in the M31 nova and switches off so quickly, despite the longer recurrence timescale.

\section*{ACKNOWLEDGEMENTS}
\label{ack}

KLP, JPO and APB acknowledge funding from the UK Space Agency. SS acknowledges partial support to ASU from NSF and NASA grants.
We thank the {\em Swift} PI, Neil Gehrels, together with the Observatory Duty Scientists and Science Planners, especially for their work in scheduling the grism observations. We thank Hans Krimm for adding V745~Sco to the BAT Transient Monitor and Brad Schaefer for useful discussion.

\bsp	
\label{lastpage}
\end{document}